\documentclass[
	a4paper, 
	10pt, 
	unnumberedsections, 
	twoside, 
]{LTJournalArticle}

\runninghead{} 

\footertext{\textit{RISC-V Summit Europe, Bologna, 8-12th June 2026}} 

\title{Monte Cimone v3: Where RISC-V Stands in High-Performance Computing} 

\author{%
	Emanuele Venieri\textsuperscript{1}\thanks{Corresponding author: \href{mailto:emanuele.venieri2@unibo.it}{\tt emanuele.venieri2@unibo.it}}, Simone Manoni\textsuperscript{1}, Giacomo Madella\textsuperscript{1}, Federico Proverbio\textsuperscript{3}, \\ Federico Ficarelli\textsuperscript{4}, Luca Benini\textsuperscript{1,2} and Andrea Bartolini\textsuperscript{1}
}

\date{\footnotesize\textsuperscript{\textbf{1}}DEI University of Bologna, Italy; \textsuperscript{\textbf{2}}ETH Zurich, Switzerland; \textsuperscript{\textbf{3}}E4 Computer Engineering Spa, Italy; \textsuperscript{\textbf{4}}CINECA, Italy}

\renewcommand{\maketitlehookd}{%
	\begin{abstract}
The Monte Cimone project provides a RISC-V testbed for High-Performacne Computing cluster. This paper presents Monte Cimone v3 (MCv3), the third iteration of the Monte Cimone RISC-V HPC cluster, integrating the SOPHGO Sophon SG2044 processor, an evolution of the SG2042 used in MCv2. We characterize MCv3 using HPL and STREAM benchmarks coupled with power measurements, and compare it against two reference platforms: the Intel Xeon Platinum 8480+~(Sapphire Rapids) and the NVIDIA Grace CPU Superchip. Our results show that the SG2044 more than doubles single-core performance and improves scalability compared to SG2042. MCv3 achieves an energy efficiency of 3.08GFLOPs/W which improves of 10x w.r.t. MCv1 and is in the range of x86-64 and Arm servers. On pure performance when normalized on the SIMD/Vector length MCv3 on its peak efficiency point (16 cores) achieves $46\%$ performance of Intel Sapphire Rapids server and $91\%$ performance of NVIDIA Grace CPU superchip. 
\end{abstract}
}

\begin{document}

\maketitle 


\section{Monte Cimone v3}

While RISC-V has gained traction in embedded and edge computing, its adoption in HPC data centers remains episodic (MCv2 and ExCALIBUR\footnote{https://riscv.epcc.ed.ac.uk/}). The Monte Cimone project evaluates the readiness of RISC-V processors for HPC workloads, aiming to establish them as a competitive alternative~\cite{mcv1}. MCv2, based on the SOPHGO SG2042, increased single-node HPL performance by $127\times$ and STREAM bandwidth by $42\times$ compared to MCv1~\cite{mcv2}. 

\begin{figure}[h]
    \centering
    \includegraphics[width=0.7\linewidth]{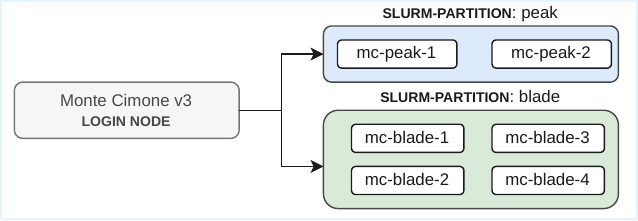}
    \caption{Comprehensive view of Monte Cimone v3. The SLURM partition \textit{Peak} includes the two SG2044 nodes, while the \textit{Blade} partition refers to the MCv2 compute nodes based on SG2042}
    \label{fig:mcv3}
\vspace{-3mm}
\end{figure}

The Monte Cimone v3 (MCv3 partition) is composed of two (SR3-RA-J-2044) compute nodes each with a Sophgo SG2044 processor. The SG2044 features 32 LPDDR5X memory channels and RVV1.0 vector support which were pain points of the MCv2 compute nodes.
 The nodes are connected to the existing cluster network as a new partition and managed through the SLURM scheduler, sharing the same software environment as the rest of the cluster, including the SPACK-based modules stack and NFS-exported user directories. An overview of the cluster configuration is shown in Figure~\ref{fig:mcv3}.

We evaluated the SG2044 nodes using the STREAM and HPL benchmarks with power measurements, and compared against two contemporary HPC platforms: a dual-socket Intel Xeon Platinum 8480+ server (Sapphire Rapids, Intel SR) and an NVIDIA Grace CPU Superchip system (NVIDIA GS). Table~\ref{tab:systems} summarizes the main hardware characteristics of the evaluated platforms.
\begin{table}[h]
\centering
\scriptsize
\begin{tabular}{lccc}
\hline
 & SG2044 & Intel SR & NVIDIA GS \\
\hline
ISA & RISC-V & x86-64 & Armv9 \\
Cores / Node & 64 & 112 (2×56) & 144 (2×72) \\
Vector ISA & RVV 1.0 & AVX-512 & SVE2 \\
Vector length & 128 bits & 2$\times$512 bits & 4$\times$128 bits \\
Frequency [GHz] & 2.6 & 2.0 & 3.1\\
Memory channels & 32 & 16 & 32\\
Memory Type & LPDDR5X & DDR5 & LPDDR5X \\
Memory & 128 GB & 2 TB & 960 GB \\
\hline
\end{tabular}
\caption{Hardware characteristics of the evaluated platforms.}
\label{tab:systems}
\vspace{-5mm}
\end{table}

\subsubsection{Benchmark setup and measurements}
Single-node performance and power consumption were evaluated using the STREAM and HPL benchmarks. These were compiled with GCC~14.2 using \texttt{-O3} and architecture-specific targets to ensure a consistent comparison. OpenBLAS~0.3.29 was used on all platforms, as no vendor-optimized BLAS library is currently available for SG2044. Although vendor libraries on Intel and NVIDIA systems may provide slightly higher HPL performance, we grounded the comparison on the same software version.

Power measurements were collected through the Intelligent Platform Management Interface (IPMI), providing platform-level power estimates across all systems, only MCv1 reported measureemnt were based on board instrumentation.
\vspace{-3mm}
\section{Experimental Results}
\subsubsection{STREAM} 
\begin{figure}[h]
    \centering
    \includegraphics[width=1.02\linewidth]{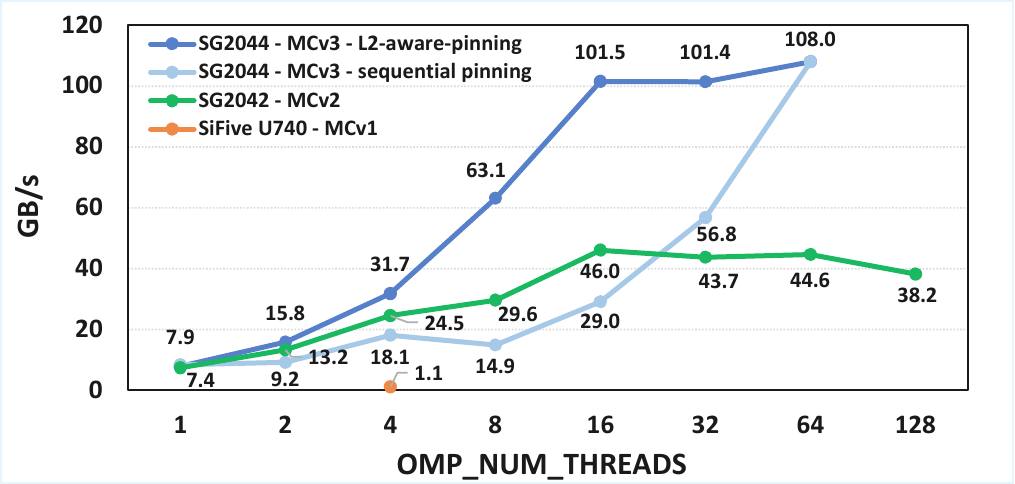}
    \caption{STREAM Triad bandwidth scaling on SG2044 with different OpenMP thread pinning strategies compared against MCv2 and Mcv1 nodes.}
    \label{fig:pinning_stream}
    \vspace{-3mm}
\end{figure}

Sequential OpenMP thread pinning on SG2044 confirms the performance characteristics previously reported in public results~\cite{nb_sg2044}. However, an L2-cache-aware pinning strategy shows that most of the redesigned memory subsystem bandwidth can already be achieved with only 16 OpenMP threads (Figure~\ref{fig:pinning_stream}). Overall, SG2044 reaches a peak bandwidth about $2.6\times$ higher than MCv2 and $100\times$ higher than MCv1.

\begin{figure}[h]
    \centering
    \includegraphics[width=\linewidth]{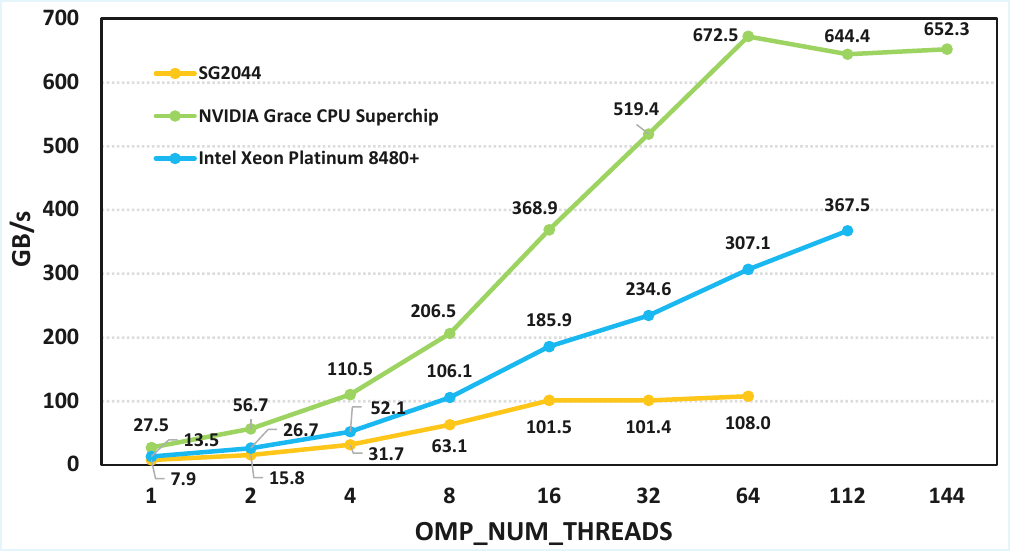}
    \caption{Cross-architecture comparison of STREAM Triad bandwidth, scaling with the number of OpenMP threads.}
    \label{fig:stream_comp}
    \vspace{-3mm}
\end{figure}

While remaining in the same order of magnitude, the Intel and NVIDIA systems achieve, at 16 threads (peak efficiency of MCv3), approximately $1.83\times$ and $3.63\times$ higher bandwidth, respectively, than the MCv3 SG2044 node. At 64 threads, the bandwidth gain over Mcv3 goes up to $2.84\times$ and $6.23\times$. Bandwidth scaling with the number of threads is shown in Figure~\ref{fig:stream_comp}. 

\vspace{-3mm}
\subsubsection{HPL} 
\begin{figure}[h]
    \centering
    \includegraphics[width=0.95\linewidth]{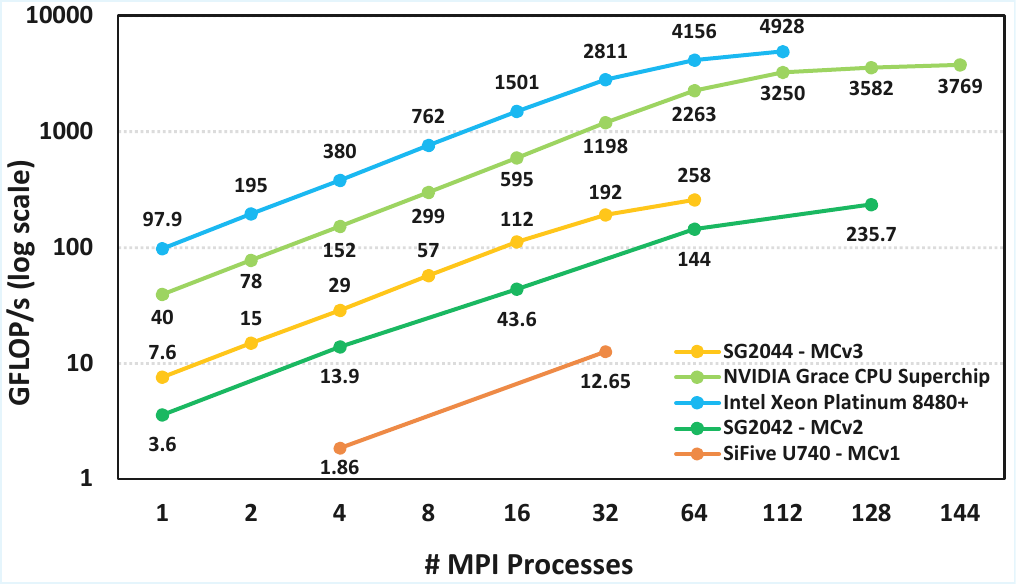}
    \caption{HPL performance comparison scaling with the number of MPI processes.}
    \label{fig:hpl}
    \vspace{-3mm}
\end{figure}

Figure~\ref{fig:hpl} compares MCv1, MCv2, and the new MCv3 platform against leading class Intel and Arm systems. Compared to SG2042, each SG2044 core achieves more than twice the performance and improved scalability up to 16 cores, likely due to the redesigned memory subsystem. The MCv3 node outperforms the MCv2 node despite using half as many cores and achieves a $139\times$ improvement over MCv1. 
Intel and NVIDIA platforms reach significantly higher absolute performance per core (NVIDIA $5.3\times$, Intel $12.9\times$), as expected given their wider vector units. However, when normalized by vector width and clock frequency, the gap between MCv3 and competing platforms shrinks substantially, reducing to $1.84\times$ and $2.62\times$ at 64 cores, and at 16 cores (peak efficiency) NVIDIA Grace and Intel Sapphire Rapids server are faster only $1.11\times$ and $2.18\times$ respectively. 
\vspace{-3mm}
\subsubsection{Power efficiency} Power efficiency results are reported in Table~\ref{tab:hpl_power}. The MCv3 node achieves $68\%$ of the NVIDIA platform's efficiency and $80\%$ of the Intel platform's efficiency in terms of GFLOPs/W.

\begin{table}[h]
\centering
\footnotesize
\begin{tabular}{lcccc}
\hline
 & MCv1 & MCv3 & NVIDIA GS & Intel SR \\
\hline
Avg Power [W] & 5.9 & 83.9 & 828 & 1208 \\
HPL [GFLOPs] & 1.86 & 258.0 & 3769 & 4928 \\
GFLOPs/W & 0.31 & 3.08 & 4.55 & 4.08 \\
\hline
\end{tabular}
\caption{Single-node HPL power efficiency comparison. Average power is obtained from IPMI readings.}
\label{tab:hpl_power}
\vspace{-3mm}
\end{table}
\vspace{-3mm}
\subsubsection{Conclusions:} The Monte Cimone projects in its third iteration shows that commercially available RISC-V compute nodes are closing the gap with their competitors in the HPC segment, lacking today only wider vector units and a many-core scalable memory subsystem.
\subsubsection{Acknowledgments: }
\footnotesize
The Monte Cimone Cluster has been funded by the DARE (g.a. 101143421) project, the Italian Research Center on High Performance Computing, Big Data, and Quantum Computing, as well as the FUTURE research project.

\vspace{-5mm}

\printbibliography 


\end{document}